 \documentclass[twocol]{ametsocV5}

\usepackage{amsmath,amsfonts,amssymb,bm}
\usepackage{mathptmx}%{times}
\usepackage{newtxtext}
\usepackage{subcaption}
\usepackage{float}
\usepackage{natbib}
\usepackage{trimclip}
\usepackage{color}
\usepackage[section]{placeins}
\usepackage{upgreek}
\usepackage{relsize,exscale}
\usepackage{epsfig}
\usepackage[mathscr]{eucal}
\usepackage{wry_abbreviations7}		% in TexInputs
\usepackage{xcolor}
\usepackage{float}
\usepackage{scalerel}
\usepackage{tikz}

\newcommand{\gamE}{\gamma_{\mathrm{E}}}

\def\Dlt{\triangle}
\def\lap{\triangle}
\def\sqphi{ {\left| \phi \right|^2} }
\def\sqphin{ {\left| \phi_0 \right|^2} }
\def\Ro{ \mathrm{Ro} }
\def\Bu{ \mathrm{Bu} }

\def\com{\, ,}
\def\per{\, .}

\newcommand{\zetamin}{\zeta_{\mathrm{min}}}
\newcommand{\uw}{u'}
\newcommand{\vw}{v'}

\definecolor{green1}{rgb}{0, 0.4, 0.4}
\definecolor{magenta1}{rgb}{1, 0, 1}
\definecolor{zereshki}{rgb}{0.4, 0, 0}
\definecolor{pink1}{rgb}{1, 0.6, 1}
\definecolor{khaki}{rgb}{0.8, 0.5, 0}
\definecolor{red1}{rgb}{0.8, 0, 0}
\definecolor{cyan1}{rgb}{0, 1, 1}
\definecolor{grey1}{rgb}{0.5, .5, .5}
\definecolor{purple1}{rgb}{0.6, 0, .5}

\newcommand{\mycircle}[2][red,fill=red]{\tikz[baseline=-0.7ex]\draw[#1,radius=#2] (0,0) circle ;}%
\newcommand{\mysquare}[2][red,fill=red]{\tikz[baseline=-0.2ex]\draw[#1,radius=#2] (0,0) rectangle (#2,#2) ;}%

\title{Interaction of near-inertial waves with an anticyclonic vortex}

    \authors{Hossein A. Kafiabad\correspondingauthor{Hossein A. Kafiabad, h.kafiabad@ed.ac.uk}
 and Jacques Vanneste }

     \affiliation{\small School of Mathematics and Maxwell Institute for Mathematical Sciences, \\
University of Edinburgh, Edinburgh, UK }

    \extraauthor{William  R. Young}
    \extraaffil{\small Scripps Institution of Oceanography, University of California, San Diego, USA}

\abstract{ Anticyclonic vortices focus and trap near-inertial waves so that near-inertial energy levels are elevated within the vortex core. Some aspects of this process, including the nonlinear  modification of the vortex by the wave, are explained by the existence of trapped near-inertial eigenmodes. These vortex eigenmodes  are easily excited by an initial wave with horizontal scale much larger than that of  the vortex radius. We study this process using a wave-averaged model of near-inertial dynamics and compare its theoretical predictions with numerical solutions of the three-dimensional Boussinesq equations. In the linear approximation, the model predicts the eigenmode frequencies and spatial structures, and a near-inertial wave energy signature that is characterized by an approximately time-periodic, azimuthally invariant pattern. The wave-averaged model represents the nonlinear feedback of the waves on the vortex 
via a wave-induced contribution to the  potential vorticity that is proportional to the Laplacian of the kinetic energy density of the waves. When this is taken into account, the modal frequency is predicted to increase linearly with the energy of the initial excitation. Both linear and nonlinear predictions agree convincingly with the Boussinesq results.}
\begin{document}

%% Necessary!
\maketitle

%%%%%%%%%%%%%%%%%%%%%%%%%%%%%%%%%%%%%%%%%%%%%%%%%%%%%%%%%%%%%%%%%%%%%
% SIGNIFICANCE STATEMENT/CAPSULE SUMMARY
%%%%%%%%%%%%%%%%%%%%%%%%%%%%%%%%%%%%%%%%%%%%%%%%%%%%%%%%%%%%%%%%%%%%%
%
% If you are including an optional significance statement for a journal article or a required capsule summary for BAMS 
% (see www.ametsoc.org/ams/index.cfm/publications/authors/journal-and-bams-authors/formatting-and-manuscript-components for details), 
% please apply the necessary command as shown below:
%
% \statement
% Significance statement here.
%
% \capsule
% Capsule summary here.

%%%%%%%%%%%%%%%%%%%%%%%%%%%%%%%%%%%%%%%%%%%%%%%%%%%%%%%%%%%%%%%%%%%%%
% MAIN BODY OF PAPER
%%%%%%%%%%%%%%%%%%%%%%%%%%%%%%%%%%%%%%%%%%%%%%%%%%%%%%%%%%%%%%%%%%%%%
%
\section{Introduction} \label{sec:intro}

The trapping of near-inertial waves by anticyclonic axisymmetric  vortices is a rare and happy case in which ocean observations \citep{KST1995,elipot2010modification,JTKT2013} are in broad agreement with theory \citep{KB1998,SGLS1999,danioux2015concentration} and with numerical models \citep{LN1998,zhai2005enhanced,AY2020}. The physical process responsible for wave trapping is that the negative core vorticity extends the internal wave band to frequency slightly below the Coriolis frequency  $f$ so that  waves  with frequency less than $f$ are   trapped within the vortex \citep{Kunze1985}.

%The concentration of wave energy inside anticyclones is well-established by both observational data \citep{KS1984,elipot2010modification,JTKT2013} and numerical simulations \citep[e.g.][]{LN1998,zhai2005enhanced,danioux2008propagation}, for which several theoretical explanations have been presented \citep{Kunze1985,YBJ1997}. Using the YBJ model, \cite{SGLS1999} quantified the dynamics of inertial waves `trapped' in a borortopic vortex, by calculating the eigenmodes of their slow oscillations. We revisit this problem by comparing his findings to the numerical solutions of Boussinesq equations and presenting some complementary analytical results. We also demonstrate that the frequency of trapped oscillations scales linearly with the wave initial energy in the Boussinesq simulations; a feature that cannot be explained by the YBJ model. To provide an explanation for this finding, we use a more complete model of \cite{XV2015} that accounts for the wave feedback on the vortex. 

\begin{figure*}[ht]
 \centerline{\includegraphics[width=0.9\textwidth]{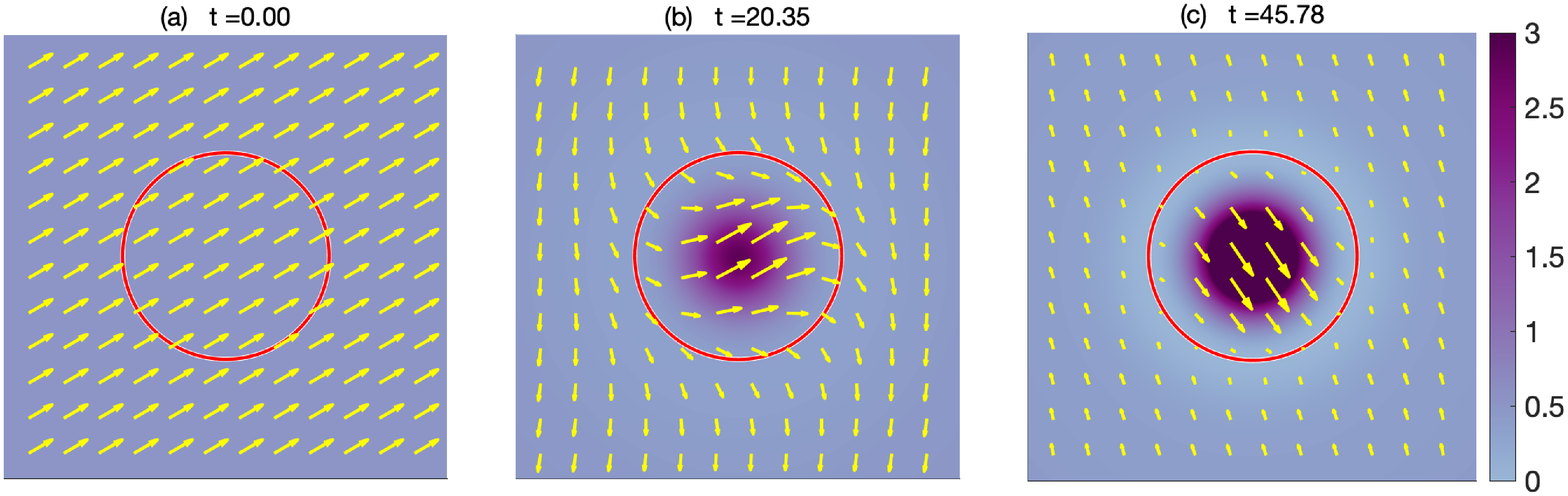}}
 \centerline{\includegraphics[width=0.9\textwidth]{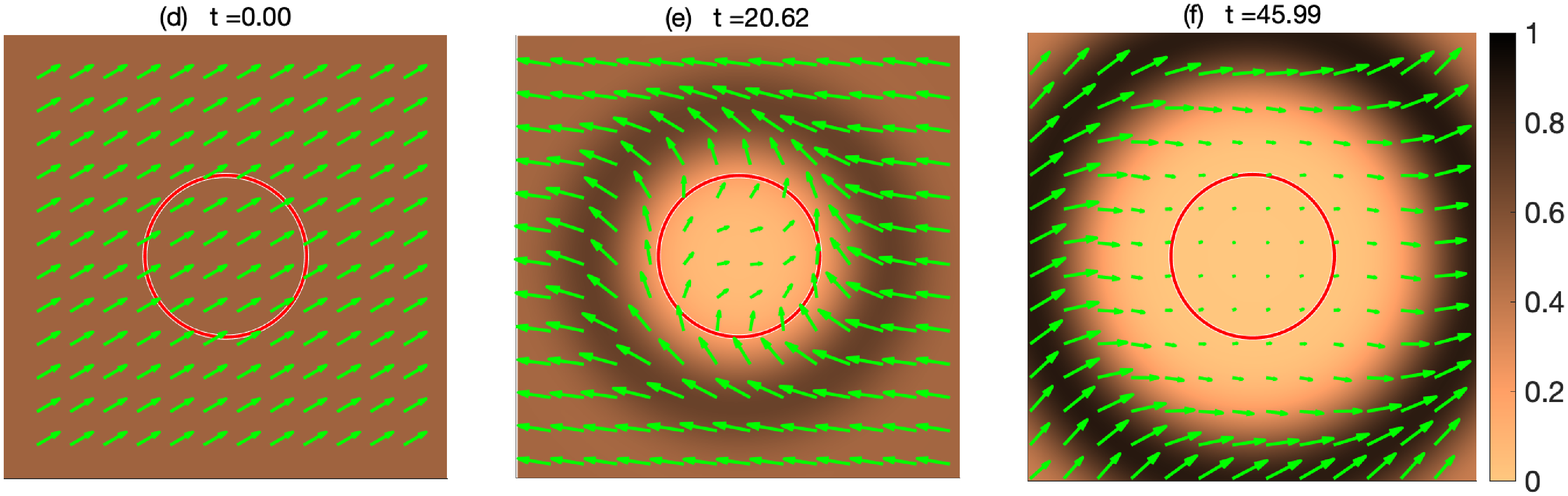}}
  \caption{Boussinesq simulation of near-inertial waves propagating on a Gaussian vortex: wave velocity vectors superimposed on the wave kinetic energy density, $(\uw^2 + \vw^2)/2$, indicated by color. The snapshots are taken at times indicated in inertial periods above each panel; the red circle has the vortex radius $a$ in \eqref{vort_profile}. The parameters are those of simulation `L13Aa' detailed in table \ref{tab:list_sim}. The upper row shows the anticyclonic case, $\ze(r)<0$; the lower row shows the cyclonic case with the same vorticity profile, $\ze(r)>0$.}\label{fig:velocity_vec}
\end{figure*}

Anticyclonic near-inertial trapping is readily illustrated with a numerical solution. The top row of figure \ref{fig:velocity_vec} shows a  solution of the Boussinesq equations  starting from  an initial condition consisting of a  barotropic  vortex superimposed with  a large-amplitude  wavy disturbance. The vortex has initial Gaussian vertical vorticity
\begin{equation}\label{vort_profile}
\zeta(x,y,z,0) = - \Ro f e^{-r^2/a^2}\com
\end{equation}
where $r=\sqrt{x^2+y^2}$ is a radial coordinate, $a$ is the vortex radius  and  $f$ is the  Coriolis parameter. The Rossby number in \eqref{vort_profile} is based on the vorticity extremum
\begin{equation}\label{defRo}
	\Ro= |\zetamin| / f\, .
\end{equation}
The vortex is distorted by a near-inertial wave that is initially horizontally uniform and vertically planar, as specified by the initial horizontal velocity
%\begin{equation}\label{initWave}
%	  \uw(x,y,z,0) + \ii \vw(x,y,z,0) =\phi_0 e^{\ii m z}\, ,
%\end{equation}
\begin{align} \label{initWave}
	  \uw&(x,y,z,0) + \ii \vw(x,y,z,0) = \phi_0 \, e^{\ii m z}\, ,
\end{align} 
where $\phi_0$ is a constant initial amplitude --- see figure \ref{fig:velocity_vec}(a) ---  and $m$ is a vertical wavenumber. In \eqref{initWave}, the  primes indicate the near-inertial-wave contribution to the velocity; this is added to the velocity associated with the vorticity $\zeta$ of the axisymmetric vortex in \eqref{vort_profile}. The initial condition has no vertical velocity and no buoyancy perturbations to the uniform buoyancy frequency $N$. If there is no vortex ($\Ro=0$) then  the disturbance in \eqref{initWave} evolves as a horizontally uniform vertical plane wave with  $\exp(\ii m z - \ii ft)$. The Gaussian  vorticity, however, perturbs the effective inertial frequency  so that the velocity vectors in figure \ref{fig:velocity_vec}(b) and (c) rotate at different rates. This de-phasing is accompanied by a concentration of wave energy into the  core of the anticyclonic vortex. For comparison, the lower row of figure \ref{fig:velocity_vec} shows the evolution of the initial disturbance in \eqref{initWave} if the sign of the vorticity in \eqref{vort_profile} is reversed so that the wave is de-phased by a cyclone:   wave energy is expelled from the cyclone.

\begin{figure*}
 \centerline{\includegraphics[width=0.95\textwidth]{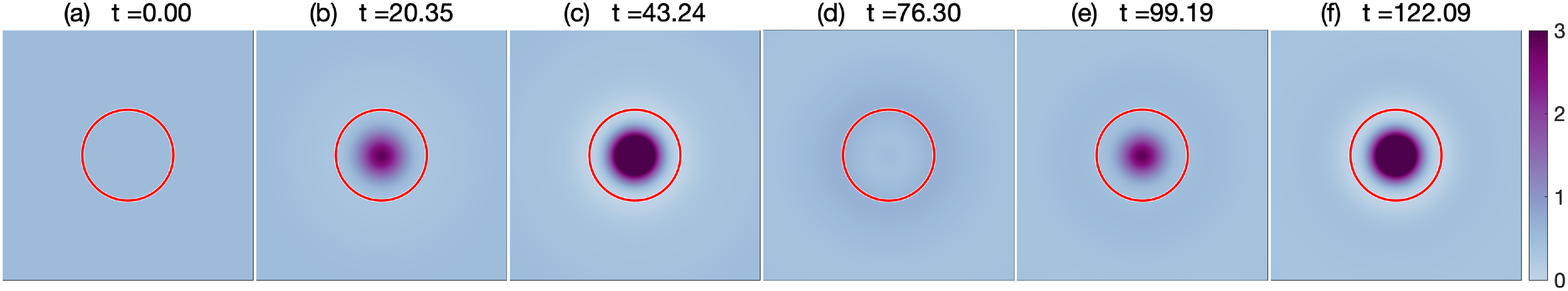}}
 \centerline{\includegraphics[width=0.95\textwidth]{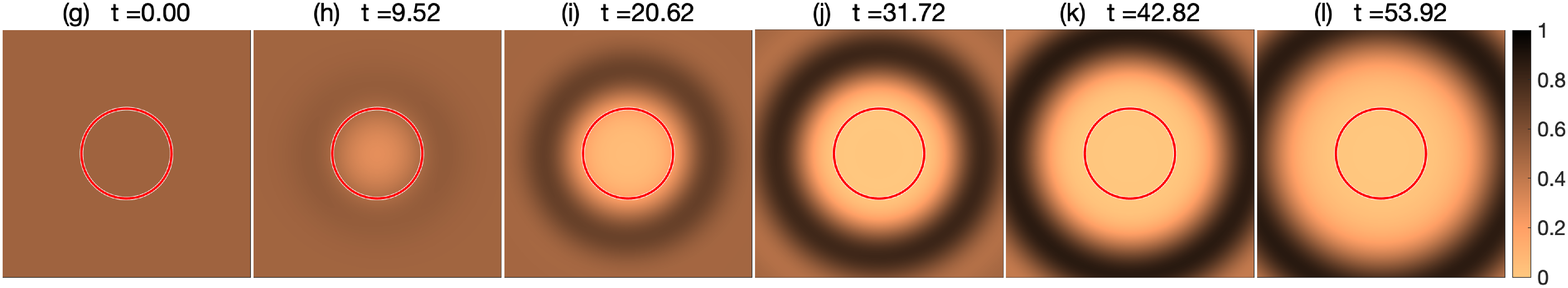}}
  \caption{Horizontal slices of wave kinetic energy $(\uw^2 + \vw^2)/2$ for the same simulation as in figure \ref{fig:velocity_vec}. Snapshots are taken at times indicated in inertial periods above each panel. The upper row shows the anticyclonic case, $\ze(r)<0$; the lower row shows the cyclonic case, $\ze(r)>0$.}\label{fig:HorizontalSlices}
\end{figure*}

The main features of the spatial pattern of  phase changes in the top row of  figure \ref{fig:velocity_vec}, and the concentration of wave energy into the anticyclonic core, can be understood  by linearizing the Boussinesq equations around a basic state consisting of  an anticyclonic  barotropic vortex, for example the Gaussian vortex  in \eqref{vort_profile},  and then solving an eigenvalue problem  to obtain the trapped near-inertial modes of the vortex \citep{KST1995,KB1998}. Instead of linearizing the Boussinesq equations, \cite{SGLS1999} used the phase-averaged equation  of Young \& Ben Jelloul (1997, YBJ hereafter) to show how the spatially uniform   initial wave in \eqref{initWave}  excites the linear eigenmodes of the vortex. The details of this linear eigenproblem problem  are, however, not without controversy and novelty:  some authors argue that the  lowest frequency of the internal wave band  is $f + \zetamin$ \citep{KB1998,JTKT2013}, while others maintain it is $f+\zetamin/2$ \citep{SGLS1999,CFA2012}. We have more to say about this issue later: we show that the lowest possible frequency of the trapped eigenmode  is $f+\zetamin/2$.

\begin{figure}
 \centerline{\includegraphics[width=0.45 \textwidth]{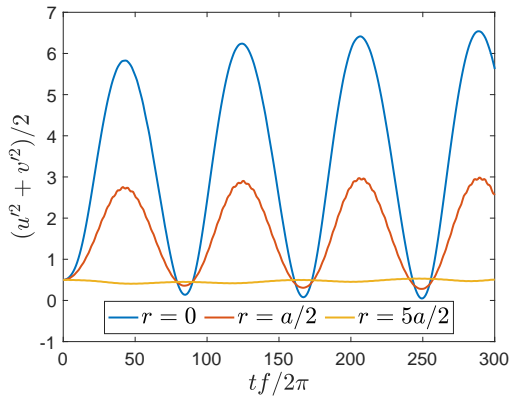}}
  \caption{Time series of the wave kinetic energy at three points given in the legend. The vortex center is at $r=0$. The parameters are those of simulation `L13Ba' detailed in table \ref{tab:list_sim}}.\label{fig:timeseries3points}
\end{figure}

The top row of  figure \ref{fig:velocity_vec}  shows that despite  the azimuthal symmetry of the base-state vortex,  the trapped eigenmode is not a radial pulsation for  which the wave velocity would have a dominant radial component. Observations of trapped near-inertial disturbances  in a warm-core ring describe a similar  structure: see  figure 14 of \cite{KST1995} and the associated  discussion. Describing  the phase of the back-rotated velocity $(u' + \ii v') \exp(\ii f t)$, \cite{KST1995} stress the ``lack of horizontal phase progression in the ring core''; this uniformity of phase within the vortex core  is a good approximation in figure \ref{fig:velocity_vec}(b) and   is strikingly appropriate    in figure \ref{fig:velocity_vec}(c).

\nocite{YBJ1997}

Further details of the initial value problem are shown in  figure \ref{fig:HorizontalSlices}. In the anticyclonic case (top row)  the initially-uniform wave kinetic energy, $(\uw^2+\vw^2)/2$, localizes inside the  vortex core  and then spreads radially  to reform an almost horizontally uniform field. This cycle of radial contraction and expansion, also evident in the time series in figure \ref{fig:timeseries3points}, repeats with a period that  is much longer than the inertial period. This sub-inertial oscillation is a signature of the vortex eigenmode and is the topic of this paper. We contrast this periodic behaviour with that obtained in a cyclonic vortex, illustrated by the bottom rows of figures \ref{fig:velocity_vec} and \ref{fig:HorizontalSlices}.  In the cyclonic case, wave kinetic energy is expelled from the vortex core, creating a void  that expands outwards in time; there is no subinertial pulsation of  wave energy.

This paper has two main aims. First, we assess how the predictions for the dynamics of trapped modes made by \cite{SGLS1999} using linear theory and the YBJ model apply to nonlinear three-dimensional Boussinesq simulations; second, we examine how nonlinear effects, specifically those associated with wave-induced changes in the vortex, impact this dynamics.

We start by formulating the vortex eigenmode problem in the YBJ approximation, focussing on the mode with azimuthally uniform backrotated velocity observed in  figures~\ref{fig:velocity_vec} and \ref{fig:HorizontalSlices} (section \ref{sec:mathformulation}). 
We add to \citeauthor{SGLS1999}'s \citeyearpar{SGLS1999} analysis by: \textit{(i)} deriving an approximation for the modal frequency in the limit of small frequency corresponding to weakly trapped modes, which gives us a handle on the number of branches of the dispersion relation; and
\textit{(ii)} showing that the lowest accessible frequency is $f+\zetamin/2$. We compare the theoretical predictions of the eigenmode problem with a series of high-resolution Boussinesq simulations (section \ref{sec:BoussinesqSim}) covering a broad range of parameters, finding an excellent agreement in spite of the complexities introduced by the excitation of a continuous spectrum of (non-trapped) modes, finite Rossby and Burger numbers, finite domain size, and  nonlinearity. We consider the effect of weak nonlinearity in section \ref{sec:FreqShift}: using the nonlinear, phase-averaged model of  \cite{XV2015}, in which the YBJ equation is coupled to a quasigeostrophic model, we predict a nonlinear frequency shift that increases the period of trapped mode and we test this predictions against Boussinesq simulations.
This quantitative comparison is a significant test of the phase-averaged  model and essential in developing confidence in its accuracy in more complicated situations, such as the propagation of near-inertial waves through  geostrophic turbulence characterized by  a  population of coherent almost axisymmetric vortices \citep{rocha2018stimulated,AY2020}.

\section{Eigenproblem for the anticyclonic vortex}\label{sec:mathformulation}
  
\subsection{YBJ vortex eigenmode problem}

Following  \cite{SGLS1999}  we use the YBJ phase-averaged description of sub-inertial evolution to solve the vortex eigenmode problem. This assumes a weak vortex, with $\Ro \ll 1$, and near-inertial wave frequencies.
Other authors have approached this same problem by linearization of the full Boussinesq equations of motion \citep{KST1995,KB1998}. This  direct assault leads to an intricate eigenproblem which reduces to the simpler YBJ eigenproblem in the relevant limit.

For the vertical-plane wave initial condition  \eqref{initWave}, the master variable used in the YBJ equation is the back-rotated velocity
\begin{equation}\label{defPhi}
	 \phi(x,y,t) = \left[ \uw(x,y,z,t) +i \vw(x,y,z,t)\right]  e^{\ii (ft - m z)},
\end{equation} 
where $\uw$ and $\vw$ are the horizontal wave velocities. Because  the vortex is barotropic, and because the waves have the  special initial condition in \eqref{initWave}, the back-rotated velocity  
$\phi$ is  independent of $z$. To a good approximation the Boussinesq solutions also have this simple structure. This  enables convenient separation of  the wave quantities from the balanced flow: the balanced component of the solutions is obtained by a vertical average. The remaining  baroclinic part of any field is  a good approximation for the wave part of that field.

Using \eqref{defPhi}, the YBJ model can be simplified for barotropic flows and constant buoyancy frequency $N$ to
\beq
\label{WV_YBJ}
	\frac{\partial \phi}{\partial t} + J(\psi, \phi) + \tfrac{\ii}{2} \ze\phi  =  \tfrac{\ii}{2} \hbar \ \Dlt \phi \com 
\eeq
where $\lap =\p_x^2+\p_y^2$ is the horizontal Laplacian. The second and third terms in \eqref{WV_YBJ}  are advection by the streamfunction $\psi$ and refraction by the vorticity $\ze = \lap \psi$ of the balanced flow. In the dispersive term on the right hand side of \eqref{WV_YBJ},
 \beq
  \hbar \defn  N^2/ (f m^2) 
\eeq 
  is the \emph{dispersivity} of near-inertial waves with vertical wavenumber $m$ (see \cite{danioux2015concentration} for further discussion on this parameter). 

Following \cite{SGLS1999} we look for eigensolutions of \eqref{WV_YBJ} in the form of
\beq \label{SGLS7}
\phi(r,\theta,t) = A(\eta) \ee^{\ii(\nu \theta -\omega t)}  \com 
\eeq
where $\eta = r / a = \sqrt{x^2+y^2}/a$ is a non-dimensional radial coordinate, $\theta$ the azimuthal angle, $\nu=0,\, 1, \ldots$ the azimuthal wavenumber, and $\omega$ the frequency of the eigenmode. Introducing \eqref{SGLS7} into \eqref{WV_YBJ} and using the Gaussian form \eqref{vort_profile} of the vortex leads to
\begin{align}\label{nondimevalue}
A_{\eta\eta} + \frac{1}{\eta} A_{\eta}+ \lambda &\left(\ee^{-\eta^2} +  \nu \frac{1-\ee^{-\eta^2}}{\eta^2} \right)A  \nonumber \\
&- \left( \sigma + \frac{\nu^2}{\eta^2} \right) A =0\com 
\end{align}
where
\begin{equation}\label{defSig}
	\sigma = - \frac{2 a^2 \omega }{\hbar}> 0
\end{equation}
is a convenient non-dimensional frequency. 
In \eqref{nondimevalue}, the strength of the vortex is characterized by  the ratio of the vortex angular momentum to the wave dispersivity 
\beq\label{lamDef}
	\lambda = \frac{a^2 |\zetamin|}{\hbar}\per
\eeq
Introducing  the Burger number 
\beq
\Bu = \left(\frac{N}{f ma } \right)^2 = \frac{\hbar}{fa^2}\com
\eeq
the vortex-strength parameter can be rewritten as the ratio
\beq
\lambda = \frac{\Ro}{\Bu}\per
\eeq 
The YBJ model assumes that $\lambda$ is fixed as $\Ro$ and $\Bu \to 0$.

 To ensure that the mode decays exponentially at great distances from the vortex center, the frequency $\sigma$ in \eqref{defSig} must be  positive so that
\beq\label{expDecay3}
A \sim \ee^{-\sqrt{\sigma} \eta} \to 0\com  \quad \textrm{as} \ \  \eta\to \infty\per
\eeq
The other boundary condition defining the eigenproblem for $\sigma$ and $A$ is that the mode has no singularity at $\eta =0$, which is equivalent to  $A'(0) = 0$.

\subsection{Azimuthal wavenumber $\nu = 0$}

In the remainder of the paper we focus on modes with $\nu=0$, which reduces the eigenproblem \eqref{nondimevalue} to
\beq\label{eigprob}
\frac{d^2 A}{d \eta^2} + \frac{1}{\eta} \frac{d A}{d \eta} + \left(\lambda e^{-\eta^2} -\sigma \right)  A = 0\per
\eeq
There are several reasons for considering only this case.
\cite{SGLS1999}  showed that trapped modes with $\nu<0$ do not exist. And, after  a vain numerical search for modes with $\nu>0$, he concluded that ``we do not know if such solutions exist, nor can we prove that they do not exist.'' We are pleased to  ignore  this open problem because the Boussinesq solution in the top row of  figures \ref{fig:velocity_vec} and \ref{fig:HorizontalSlices}  shows that the initial condition in figure \ref{fig:velocity_vec}(a) excites only  $\nu=0$ modes. With $\nu=0$, the eigenproblem in \eqref{eigprob} is the same as Schr\"odinger's equation with a two-dimensional Gaussian potential.

The absence of modes with non-zero $\nu$ in figures \ref{fig:velocity_vec} and \ref{fig:HorizontalSlices}  is remarkable because the initial condition breaks  azimuthal symmetry by selecting a special direction: all the velocity vectors in figure  \ref{fig:velocity_vec}(a) point North-East. Despite this   broken azimuthal symmetry,  the  trapped disturbance is axisymmetric in the 
sense\footnote{The vector field $(u,v)$ is not axisymmetric in the usual sense, that is, $\theta$-independent and pointing in the radial direction.}
that:  (a) velocity vectors lying on any circle of radius $r$ in top row of figure \ref{fig:velocity_vec}(a) are identical to one another; and (b) the kinetic energy  density  in figure \ref{fig:HorizontalSlices} is axisymmetric. As discussed in section \ref{sec:intro}, this is consistent with the structure observed by \cite{KST1995} in a warm-core ring.

In the Boussinesq eigenproblem of \cite{KST1995} and  \cite{KB1998} the master variable is the radial component of velocity, $u_r(r,\theta,t)$. The relation $u_r + \ii u_\theta = (u + \ii v) e^{-\ii \theta}$ then shows that a $\theta$-independent backrotated velocity $\phi$, i.e. our $\nu=0$, corresponds to the azimuthal wavenumber  $n=-1$ of  \cite{KB1998}. While  use of the back-rotated velocity $\phi$ in a problem with axial symmetry might at first sight  seem unnatural, the simplicity of the YBJ equation and associated eigenproblem shows its effectiveness in examining  near-inertial waves in small-Rossby-number flows; see \cite{SGLS1999} for a detailed discussion.

\subsection{Solution of the eigenproblem \eqref{eigprob}}\label{sec:eigsolution}

\begin{figure}
 \centerline{\includegraphics[width=19pc]{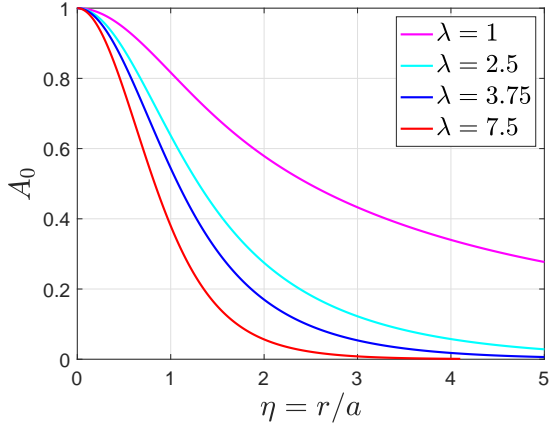}}
 \caption{Eigenfunctions corresponding to the largest eigenvalue $\sigma_0$ for different $\lambda$. This is the zeroth eigenmode, which is characterized by having no zeros. As the vortex strength $\lambda$ increases, the mode becomes more tightly trapped to the vortex core and its eigenvalue, $\sigma_0$,  increases: see figure \ref{fig:twobranches} for $\lambda$ as a function of $\sigma$.}\label{fig:Eigfun4largestsigma}
\end{figure}

\begin{figure} 
   \centerline{\includegraphics[width=19pc]{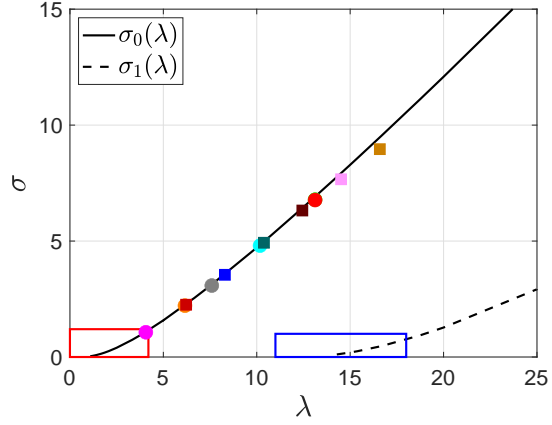}}
  \caption{Non-dimensional frequency $\sigma$ of trapped modes for different $\lambda$ derived by numerical solution of  the YBJ eigenproblem \eqref{eigprob}. The two branches above correspond to the zeroth and first modes. The coloured circles correspond to frequencies for values of $\lambda$ used for the simulations in table \ref{tab:list_sim}. See figure~\ref{fig:magnifiedbranches} for a magnified view of the two rectangles.}\label{fig:twobranches}    
\end{figure}

\begin{figure}
 \centerline{\includegraphics[width=19pc]{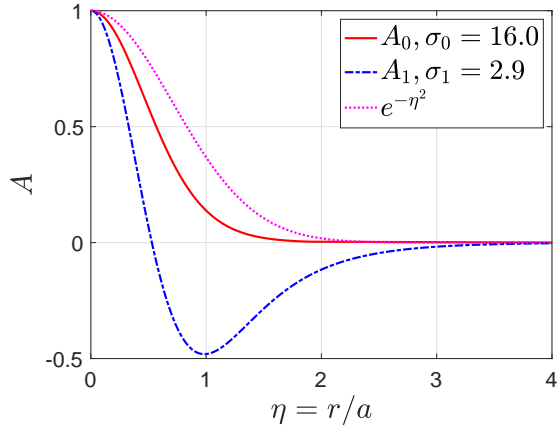}}
  \caption{Two eigenmodes  of the Gaussian vortex with  $\lambda = 25$ with their corresponding eigenvalues  in the legend. The Gaussian $\exp(-\eta^2)$ is shown  as  guidance.  The zeroth mode corresponds to $\sigma_0 = 16.0$; the first mode, with $\sigma_1= 2.9$ and one zero at $\eta = 0.53$, is more weakly trapped than the zeroth mode.} \label{fig:twoEigFuns}
\end{figure}

We now turn to solution of the boundary-value problem in  \eqref{expDecay3} and \eqref{eigprob}. Asymptotic calculations detailed in Appendix 
A show that for all values of $\lambda$,  including very weak vortices with $\lambda \ll 1$,  there is at least one trapped mode. We refer to this important solution as the zeroth mode and denote its corresponding  eigenfunction and eigenvalue  by $A_0$ and $\sigma_0$, respectively: numerical results in  figures~\ref{fig:Eigfun4largestsigma}  illustrate the form of the zeroth-mode  solution. The eigenproblem in \eqref{eigprob} is analogous to the quantum mechanical problem of trapping in a two-dimensional Gaussian potential well;  in that context the zeroth mode is known as the ground state of the well. 

As $\lambda$ increases, additional trapped modes appear through a sequence of bifurcations arising at $\lambda = \lambda_n$, $n=1$, $2$, $\cdots$.
Figure \ref{fig:twobranches}  shows the first two eigenbranches, $\sigma_0(\lambda)$ and $\sigma_1(\lambda)$. The structure of the corresponding eigenfunctions is illustrated in figure 
\ref{fig:twoEigFuns}, which shows $A_0(\eta)$ and $A_1(\eta)$ for $\lambda = 25$.

\begin{figure} 
   \centerline{\includegraphics[width=19pc]{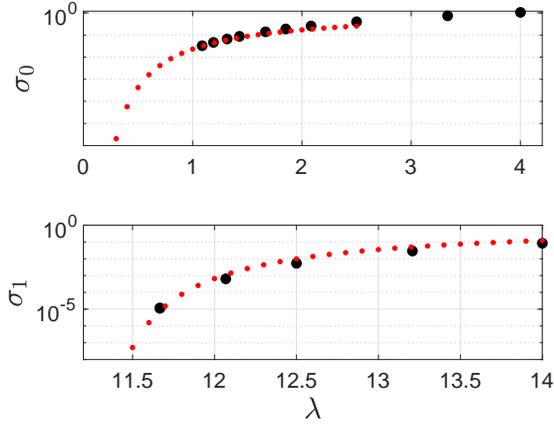}}
  \caption{Magnified view of  the red rectangle (top) and blue rectangle (bottom) from  figure \ref{fig:twobranches}, with logarithmic vertical axis. Numerical results (large black dots) are comparted with the asymptotic results  \eqref{asymptotic7} and  \eqref{eq:smallSigma} (small red dots).}\label{fig:magnifiedbranches}    
\end{figure}

The bifurcations giving rise to new branches of the dispersion relation can be analyzed by solving the eigenvalue problem in the asymptotic limit $\sigma \to 0$ corresponding to weakly trapped modes: see \eqref{expDecay3}. This is done in Appendix A where we find that the first three branches arise for
\beq
\lbrace  \lambda_0\com \lambda_1\com  \lambda_2\rbrace= \lbrace{0\com11.1\cdots \com  \, 35.1\cdots \rbrace} \per
\eeq
(The second mode is off-stage in figure \ref{fig:twobranches}.)
On each branch $\sigma \to 0$ very rapidly as $\lambda \to \lambda_n$: our analysis shows that
\beq\label{asymptotic7}
\sigma_n \sim \exp\left( - \frac{d_n}{\lambda - \lambda_n}\right)\com \qquad \text{as $\lambda \downarrow \lambda_n$}\com 
\eeq
where the $d_n$ are constant that can be evaluated explicitly. For the zeroth mode $\lambda_0=0$, and \eqref{asymptotic7} reduces to
\beq \label{eq:smallSigma}
\sigma_0 \sim e^{2 (\ln 2 - \gamE) - 4 / \lambda} \quad \textrm{as} \ \ \lambda \downarrow 0 \com 
\eeq 
where $\gamE$ is the Euler-Mascheroni constant. The exponential sensitivity of $\sigma_n(\lambda)$ to $(\lambda-\lambda_n)^{-1}$ explains the very flat curves as $\lambda\downarrow \lambda_n$ in figure \ref{fig:twobranches}.  In figure \ref{fig:magnifiedbranches} we verify the asymptotic prediction \eqref{asymptotic7} for $n=0$ and $1$ by comparison  with the numerical solutions of the eigenproblem \eqref{eigprob}.
%the asymptotic behaviour of $\sigma_0(\lambda)$ and $\sigma_1(\lambda)$ as $\lambda \downarrow \lambda_0$  and $\lambda \downarrow \lambda_1$  are magnified in logarithmic scale , where a reassuring agreement is observed.}

%For example, figure \ref{fig:twoEigFuns} shows the two eigenmodes  of a  vortex with $\lambda = 25$. In addition to the zeroth mode there is also the  first mode, with a zero at $\eta = 0.53$. 

In the Boussinesq numerical solution the initial condition in \eqref{initWave} will project onto all of the trapped eigenmodes of  the Gaussian vortex in \eqref{vort_profile}. The vortex used for illustrative purposes in figures \ref{fig:velocity_vec} through \ref{fig:timeseries3points}  has $\lambda = 13.1$. Because  
\beq 
11.1 <13.1 < 35.1\com
\eeq
 this vortex  has two trapped modes (the zeroth and first). But only the zeroth mode  is evident in figures \ref{fig:velocity_vec} through \ref{fig:timeseries3points}, presumably because the initial condition projects only weakly on the first mode.
%
%\textcolor{red}{[Do we believe the above? I'd feel better if we could say that the first mode also has a much weaker projection onto the initial condition than the zeroth mode. I suspect this is why we never see the first mode.]} \HAK{I prefer so too. I think the first mode DOES have a weak projection. The results of table 1 for values less than 11.1 match $\sigma_0$ better.}

\subsection{The lowest vortex-mode frequency \label{lowFreak}}

A bound on the frequency of sub-inertial oscillations can be obtained from the eigenproblem \eqref{eigprob}.
Untangling the non-dimensionalization, the total frequency of the  eigenmode in dimensional variables is
\beq
f+\Bu \, f \omega = f\big(1 - \half \Bu \sigma \big)= f + \half \frac{\sigma}{\lambda} \, \zetamin\per
\label{lowFreak3}
\eeq
For the $\nu=0$ modes studied here, the issue of whether the lowest frequency of the internal wave band  is $f+\zetamin$ or $f+\zetamin/2$ devolves to whether the ratio $\sigma/\lambda$ in \eqref{lowFreak3} is ever greater than $1$.  Examination of figure \ref{fig:twobranches} indicates that for the Gaussian vortex $\sigma/\lambda$  is less than one and thus for these modes $f+\zetamin/2$ is the lowest possible frequency. 

We now establish this property for a general compact vortex, with a vorticity profile $\zeta(r)$ satisfying 
\beq
\zeta(0)=\zetamin\leq \zeta(r)<0\per
\eeq
The generalisation of the eigenproblem \eqref{eigprob} is
\beq\label{eigprobgen}
\frac{d^2 A}{d \eta^2} + \frac{1}{\eta} \frac{d A}{d \eta} + \left(\lambda f(\eta) -\sigma \right)  A = 0.
\eeq
where $0< f(\eta) \le 1$ is (minus) the non-dimensional vorticity profile. Multiplying by $\eta$, integrating and using the boundary conditions of trapped modes leads to
\beq
\frac{\sigma}{\lambda} = \frac{\int_0^\infty f(\eta) A(\eta) \eta \, \dd \eta}{\int_0^\infty  A(\eta) \eta \, \dd \eta}.
\eeq
The zeroth mode, also known as the ground state, has the largest frequency $\sigma$ and a sign-definite eigenfunction, hence
\beq
\frac{\sigma_n}{\lambda} \le \frac{\sigma_0}{\lambda} \le \max_\eta f(\eta) = 1,
\eeq
which completes the argument.

%The eigenproblem obtained via substitution of $\phi = A(r) \exp(- \ii \omega t)$ into \eqref{WV_YBJ} is
%\beq
%\half \hbar \left(A''+ r^{-1} A' \right) + \left( \omega - \half  \zeta\right) A=0\com 
%\label{lowFreak7}
%\eeq
%where the prime denotes $\p_r$. In \eqref{lowFreak7} both $\zeta(r)$ and $\omega$ are negative. Might  $\omega$ might be less than $\zetamin/2$? The answer is no: $\omega$ must be greater than  $\zetamin/2$.
%
%To prove this result first multiply  \eqref{lowFreak7} by $r$, and then integrate  from $r=0$ to $\infty$;  confining attention to trapped modes with $A(r \to \infty)=0$ one finds that
%\beq
%\int_0^\infty\!\! r \left( \omega - \half  \zeta\right) A\, \dd r =0\per
%\label{lowFreak11}
%\eeq
%The zeroth mode, $A_0(r)$, has no zeros at any $r$, and the corresponding eigenfrequency, $\omega_0$,  is at  the bottom of the  spectrum:
%\beq
%\omega_0<\omega_1 < \omega_2 < \cdots < \omega_n  <0\com
%\eeq
%where $n(\lambda)$ is the number of discrete modes of the vortex.
%But the integrand on the left of \eqref{lowFreak11} must have a zero somewhere, and for the zeroth mode this can only be a zero of 
%\beq
% \omega_0 - \half \zeta(r)\per
% \label{lowFreak17}
%\eeq
%Now as $r \to \infty$, the function above limits to $\omega_0<0$. Thus at $r=0$, the function in \eqref{lowFreak17} is positive, which requires that $\omega_0>\zetamin/2$.

\section{Comparison with numerical solutions of Boussinesq equations}\label{sec:BoussinesqSim}

We now assess the analytical results of previous sections against a suite of high-resolution non-hydrostatic Boussinesq solutions in a triply periodic domain. In these simulations, the flow is initialized with the planar wave  in \eqref{initWave} superimposed on the barotropic vortex  in \eqref{vort_profile}. To maintain the periodicity of the initial field, the Gaussian vortex in (\ref{vort_profile}) is slightly modified by discretizing it in the Fourier space and truncating the unresolved high-wavenumber modes. A de-aliased pseudospectral solver detailed in \cite{kafiabad2020wave} is used to derive the numerical solutions, and a third-order Adams--Bashforth scheme is used for time-integration. A hyperdissipation of the form $\nu_\mathrm{h} (\partial_x^2 + \partial_y^2)^4 +\nu_\mathrm{z} \partial^8_z$  is used in  the momentum and buoyancy equations. The flow parameters and  set-up  are  in table \ref{tab:list_sim}. 
 
% \begin{table}[h]
%\caption{Parameter values and numerical set-up for the main Boussinesq simulation.}\label{tab:main_sim}
%\begin{center}
%  \begin{tabular}{ | c  c  c | }
%    \hline
%    $L$ & Horizontal domain size & $2 \pi$ \\
%    $H$ & Vertical domain size & $2 \pi / 36$ \\
%    $N_x=N_y$ & Grid points on x- and y-axis & 1152 \\
%     $N_z$ & Grid points on z-axis & 96 \\
%    $f$  & Coriolis parameter & $200$ \\ 
%    $N$ & Brunt--V\"ais\"al\"a frequency & $1600$ \\ 
%    $a$  & Vortex radius & $0.45$ \\ 
%    $m$ & Vertical wavenumber & $288$ \\
%    $E_0$ & Initial wave kinetic energy & $1/2$ \\ 
%    $\Ro$ & Rossby number & $0.05$ \\
%    $\Bu$ & Burger number & $0.0038$ \\
%    $\lambda$ & Ratio of $\Ro/\Bu$ & $13.1$ \\
%    $\nu_\mathrm{h} $  & Horizontal hyperviscosity &  $5 \times 10^{-18}$ \\
%    $\nu_\mathrm{z}$   & Vertical hyperviscosity & $5 \times 10^{-23}$ \\
%    \hline
%  \end{tabular}
%\end{center}
%\end{table}

 \begin{table*}[ht]
\caption{List of Boussinesq simulations with their corresponding parameter values. All simulations share the following parameters:  horizontal domain size $L = 2 \pi$, vertical domain size  $H = 2 \pi / 36$, number of grid points on x- and y-axis $N_x=N_y=1152$, , number of grid points on z-axis $N_z = 96$, horizontal hyperviscosity $\nu_\mathrm{h} = 5 \times 10^{-18}$,  vertical hyperviscosity $\nu_\mathrm{z} = 5 \times 10^{-23}$ }\label{tab:list_sim}

\begin{center}
  \begin{tabular}{ |c  c  c  c  c  c  c  c  c  c  c  c  c  c| }
    \hline
    simulation & $f$ & $N$ & $a$ & $m$ & $\Delta t$ &$E_0$ & $E_G$ & $\Ro$ & $\Bu$ & $\lambda$ & $\sigma_\mathrm{sim}$ & $\frac{|\sigma_0 - \sigma_\mathrm{sim} |}{\sigma_0}$ & Marker \\
    %L3 & $200$ & $1600$ & $0.25$ & $288$ & $1.74 \times 10^{-4}$ & $0.5$ & $0.009$ & $0.04$ & $0.0123$ & $3.24$ & $0.65$ &  $6.85\%$ & \textcolor{cyan}{$\times$} \\
    L4 & $200$ & $1600$ & $0.28$ & $288$ & $1.74 \times 10^{-4}$ & $0.1$ & $0.014$ & $0.04$ & $0.0098$ & $4.06$ & $1.07$ &  $1.93\%$ & \mycircle[magenta1, fill=magenta1]{2.3pt} \\
    L6A & $200$ & $1300$ & $0.28$ & $288$ & $1.74 \times 10^{-4}$ & $0.05$ & $0.014$ & $0.04$ & $0.0065$ & $6.16$ & $2.21$ &  $1.28\%$ & \mycircle[orange, fill=orange]{2.3pt} \\
    L6B-R03 & $200$ & $1600$ & $0.40$ & $288$ & $1.74 \times 10^{-4}$ & $0.05$ & $0.026$ & $0.03$ & $0.0048$ & $6.22$ & $2.26$ &  $1.17\%$ & \mysquare[red1, fill=red1]{4pt} \\
    L7 & $200$ & $1600$ & $0.50$ & $180$ & $1.74 \times 10^{-4}$ & $0.1$ & $0.212$ & $0.06$ & $0.0079$ & $7.59$ & $3.08$ &  $1.48\%$ & \mycircle[grey1, fill=grey1]{2.3pt} \\
    L8a & $200$ & $960$ & $0.40$ & $288$ & $1.30 \times 10^{-4}$ & $0.5$ & $0.046$ & $0.04$ & $0.0048$ & $8.29$ & $3.45$ & $3.57\%$ &  \\
    %L8b & $200$ & $960$ & $0.40$ & $288$ & $1.74 \times 10^{-4}$ & $0.2$ & $0.046$ & $0.04$ & $0.0048$ & $8.29$ & $3.52$ & $1.57\%$ & \textcolor{blue}{$\times$} \\
    L8c-R04 & $200$ & $1600$ & $0.40$ & $288$ & $1.74 \times 10^{-4}$ & $0.05$ & $0.046$ & $0.04$ & $0.0048$ & $8.29$ & $3.55$ & $0.90\%$ &  \mysquare[blue, fill=blue]{4pt} \\
    L10A & $200$ & $1300$ & $0.36$ & $288$ & $1.56 \times 10^{-4}$ & $0.1$ & $0.032$ & $0.04$ & $0.0039$ & $10.18$ & $4.80$ &  $0.69\%$ & \mycircle[cyan1, fill=cyan1]{2.3pt} \\
    L10B-R05 & $200$ & $1600$ & $0.4$ & $288$ & $1.74 \times 10^{-4}$ & $0.05$ & $0.071$ & $0.05$ & $0.0048$ & $10.37$ & $4.93$ &  $0.82\%$ & \mysquare[green1, fill=green1]{4pt} \\
    L10C & $200$ & $1600$ & $0.45$ & $216$ & $1.74 \times 10^{-4}$ & $0.1$ & $0.205$ & $0.07$ & $0.0068$ & $10.33$ & $4.87$ &  $1.40\%$ & \mycircle[purple1, fill=purple1]{2.3pt}\\
    L12-R06 & $200$ & $1600$ & $0.4$ & $288$ & $1.74 \times 10^{-4}$ & $0.05$ & $0.103$ & $0.06$ & $0.0048$ & $12.44$ & $6.32$ &  $1.66\%$ & \mysquare[zereshki, fill=zereshki]{4pt} \\
    L13Aa & $200$ & $1600$ & $0.45$ & $288$ & $1.30 \times 10^{-4}$ & $0.5$ & $0.105$ & $0.05$ & $0.0038$ & $13.12$ & $6.58$ & $4.88\%$ &  \\
    L13Ab & $200$ & $1600$ & $0.45$ & $288$ & $1.30 \times 10^{-4}$ & $0.2$ & $0.105$ & $0.05$ & $0.0038$ & $13.12$ & $6.71$ & $3.00\%$ &  \\
    L13Ac & $200$ & $1600$ & $0.45$ & $288$ & $1.30 \times 10^{-4}$ & $0.1$ & $0.105$ & $0.05$ & $0.0038$ & $13.12$ & $6.76$ & $2.27\%$ &  \\
    L13Ad & $200$ & $1600$ & $0.45$ & $288$ & $1.30 \times 10^{-4}$ & $0.05$ & $0.105$ & $0.05$ & $0.0038$ & $13.12$ & $6.77$ & $2.08\%$ & \mycircle[red, fill=red]{2.3pt} \\
    %L13Ba & $200$ & $960$ & $0.27$ & $288$ & $1.74 \times 10^{-4}$ & $0.5$ & $0.019$ & $0.05$ & $0.0038$ & $13.12$ & $6.39$ & $7.50\%$ & \textcolor{green}{$\mcirc$} \\
   % L13Bb & $200$ & $960$ & $0.27$ & $288$ & $1.74 \times 10^{-4}$ & $0.2$ & $0.019$ & $0.05$ & $0.0038$ & $13.12$ & $6.63$ & $4.40\%$ & \textcolor{green}{+} \\
    %L13Bc & $200$ & $960$ & $0.27$ & $288$ & $1.74 \times 10^{-4}$ & $0.1$ & $0.019$ & $0.05$ & $0.0038$ & $13.12$ & $6.73$ & $2.60\%$ & \textcolor{green}{$\times$} \\
    L13Bd & $200$ & $960$ & $0.27$ & $288$ & $1.74 \times 10^{-4}$ & $0.05$ & $0.019$ & $0.05$ & $0.0038$ & $13.12$ & $6.78$ & $1.85\%$ &  \mysquare[green, fill=green]{4pt}  \\
    L14-R07 & $200$ & $1600$ & $0.4$ & $288$ & $1.74 \times 10^{-4}$ & $0.05$ & $0.140$ & $0.07$ & $0.0048$ & $14.51$ & $7.66$ &  $3.34\%$ & \mysquare[pink1, fill=pink1]{4pt} \\
    L16-R08 & $200$ & $1600$ & $0.4$ & $288$ & $1.74 \times 10^{-4}$ & $0.05$ & $0.182$ & $0.08$ & $0.0048$ & $16.59$ & $8.96$ &  $5.39\%$ & \mysquare[khaki, fill=khaki]{4pt} \\
    \hline
  \end{tabular}
\end{center}
\end{table*}

%\begin{figure*} [h]
%    \centering
%    \begin{minipage}{.5\linewidth}
%         \centerline{\includegraphics[trim=1cm 0 .5cm 0, width=1\linewidth]{figz/weaktrap_lam0.eps}}
%    \end{minipage}%
%    \begin{minipage}{.5\linewidth}
%          \centerline{\includegraphics[trim=1cm 0 .5cm 0, width=1\linewidth]{figz/weaktrap_lam1.eps}}
%    \end{minipage}%    
% \caption{Magnified regions of figure \ref{fig:twobranches}  for the red rectangle (left) and blue rectangle (right), where the vertical axis is scaled logarithmically to compared the results with the asymptotic limits of (\ref{asympt4lambda_n}) shown by red dots.} \label{fig:zoombranches}    
%\end{figure*}
%

The first aspect of the theoretical results that can be compared against the numerical solutions of Boussinesq equations is the frequency of sub-inertial oscillations such as those observed in figure~\ref{fig:timeseries3points} and the top row of figure~\ref{fig:HorizontalSlices}. For each simulation of table \ref{tab:list_sim}, we estimate the scaled frequency, $\sigma_\mathrm{sim}$, defined in (\ref{defSig}) by averaging the times between  consecutive troughs and peaks of wave energy at $r=0$. We also solve \eqref{eigprob} for the value of $\lambda$ in each simulation to derive the zeroth eigenfrequency, $\sigma_0$. In the second last column of table \ref{tab:list_sim}, the normalized difference between $\sigma_\mathrm{sim}$ and $\sigma_0$ is shown. Within the range of $\lambda\leq 13$, this difference remains less than $2\%$, if the simulations with the lowest wave energy level for each set of parameters are considered. This remarkable agreement is  shown  in figure~\ref{fig:twobranches} by superimposing $\sigma_\mathrm{sim}$ on the zeroth eigenbranch. The colored markers in this figure correspond to those in the last column of table  \ref{tab:list_sim}.  For $\lambda > 13$, the projection of initial condition on the first eigenvector (in addition to the zeroth one) affects the slow modulation of wave energy observed in the simulations: this first mode component increases the relative difference between $\sigma_\mathrm{sim}$ and $\sigma_0$.

\begin{figure}
    \centerline{
     \includegraphics[width=19pc]{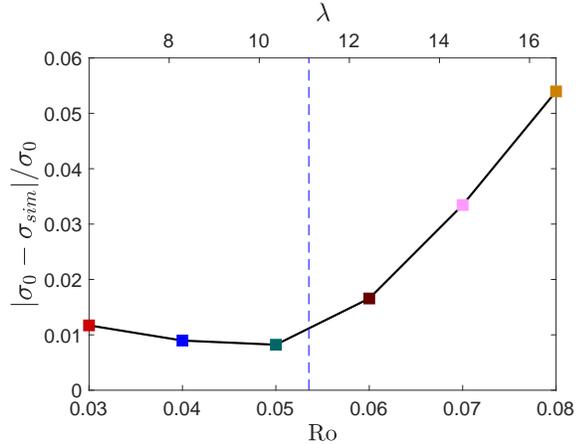}}
\caption{Relative difference between the modal frequency estimated from Boussinesq simulations and the smallest eigenfrequency of the YBJ model as function of $\Ro$. The colour-coded markers refer to the simulations in table \ref{tab:list_sim}. The second eigenmode exists to the right side of the dashed vertical line marking $\lambda=\lambda_1$.}
\label{fig:errSigma}
\end{figure}

The simulations with the same parameters, but increasing  initial wave energy, reveal a systematic dependence of the modal frequency on the amplitude of the initial wave. This dependence is not captured by the YBJ model, because it neglects the nonlinear wave feedback onto the  balanced flow. Analogy with other nonlinear oscillators suggests that this feedback likely results in a frequency shift that  depends on the wave energy level. We will discuss this phenomenon in depth in the next section and estimate the frequency shift using  the coupled model of \cite{XV2015}. Setting this frequency shift aside, the remaining differences between results based on the YBJ model and the Boussinesq solution can plausibly be attributed to some combination of:

\begin{enumerate}
\item[\textit{(i)}] inaccuracy in the YBJ equation resulting from the finite Rossby and Burger numbers;
\item[\textit{(ii)}] finite domain size of the Boussinesq code;
\item[\textit{(iii)}] low-resolution sampling frequency of the times series used to calculate $\sigma_\mathrm{sim}$.
\end{enumerate} 

Figure \ref{fig:errSigma} displays $|\sigma_0 - \sigma_\mathrm{sim} |/\sigma_0$ as function of $\Ro$ for a suite of simulations with identical parameters, but varying $\Ro$. Increasing $\Ro$ increases the discrepancy between $\sigma_0$ and $\sigma_\mathrm{sim}$. This is partly due to  nonlinear effects,  not captured in the linear YBJ model -- issue \textit{(i)} -- and partly due to excitation of the higher eigenmodes  that appear as  $\Ro$ and therefore   $\lambda$, is increased. For very small values of $\lambda$, a long integration time is required to capture a few oscillations, which leads to re-entering of the waves back to the domain and  interactions with the mean flow and other waves -- issue \textit{(ii)}. Such a long time scales, however,  do not have realistic implications in the interaction of oceanic flows with waves. For instance, the eigenperiod of the case $\lambda = 3$ is more than 230 inertial periods.

\begin{figure}
    \centerline{
     \includegraphics[width=19pc]{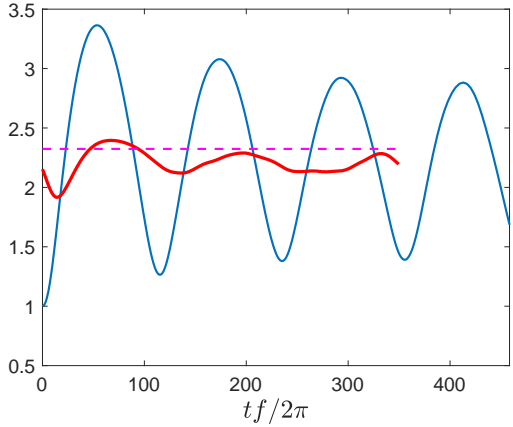}}
\caption{Wave amplitude $|\phi|$ at $r=0$ (blue curve), back-rotated velocity at $r=0$ after removing the continuum, $\left| \phi  - \phi_\mathrm{cont} \right|$ (red curve), and  $\alpha_0\phi_0=2.32$ calculated from \eqref{alpha0} (dashed magenta) for simulation `L8a'.}
\label{fig:RemoveConty}
\end{figure}

\begin{figure}
    \centerline{
     \includegraphics[width=19pc]{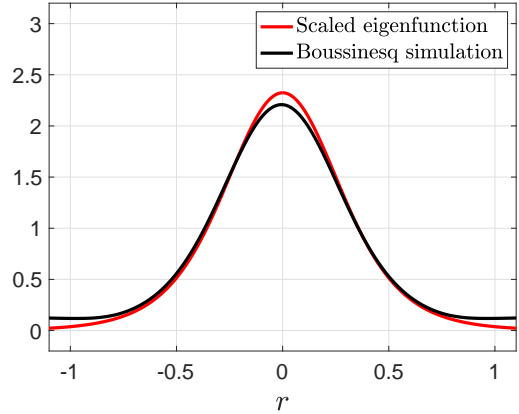}}
\caption{Scaled eigenfunction $\alpha_0 \phi_0 A_0(r)$, with $\alpha_0$ computed using (\ref{alpha0}) (red curve) compared with $|\phi(r,t) - \phi_\mathrm{cont}(r,t)|$ extracted from simulation `L8a' data (black curve).}
\label{fig:compare_eigfun}
\end{figure}

Comparing the eigenfunctions of section \ref{sec:mathformulation}.\ref{sec:eigsolution} against the simulations is less straightforward, because the  initial condition \eqref{initWave} excites not only trapped vortex eigenmodes but also a continuum spectrum \citep{SGLS1999}. Taking this into account, the solution of \eqref{WV_YBJ} can be written as
\beq \label{AllmodeConty}
\phi = \phi_0 \left[ \sum_{p=0}^{N-1} \alpha_p  A_p (r/a) \ee^{- \ii \omega_p t} + \phi_{\text{cont}}(r, \tau) \right],
\eeq 
where $\alpha_p$ is the projection of the normalized initial condition onto mode $A_p$ and  $N=N(\lambda)$ is the  number of trapped modes for given $\lambda$. Here we set $N=1$ since we are considering  values of $\lambda$ where the higher eigenmodes either do not exist or their eigenfrequency is much lower than $\omega_0$. The term $\phi_\text{cont}$ is the ``continuum remnant'' that is left over because the trapped modes do not form a complete basis; \cite{SGLS1999} shows that $\phi_\text{cont}$ depends logarithmically on time for large time. Because this time dependence is slow compared with $1/ \omega_0$, the continuum remnant can be estimated by integrating over one eigenperiod,
\beq\label{conty3}
\phi_\mathrm{cont}(r,t) \approx \frac{\omega_0}{2\pi} \int_t^{t+2\pi / \omega_0} \!\!\!\!\!\!\!\!\!\!\!\!\phi(r,s) \,  \dd s \com
\eeq
and removed from the solution to obtain
\begin{equation}\label{deriveA}
	\alpha_0 \phi_0 \, A_0(r/a)=  \phi(r/a,t)  - \phi_\mathrm{cont}(r/a,t).
\end{equation} 
$A_0$ is orthogonal to all higher modes $A_p$ ($p>0$) and to $\phi_{\text{cont}}$. Hence, after multiplying both sides of (\ref{AllmodeConty}) by $ A_0 \eta$ and integrating (at $t=0$), $\alpha_0$ is 
\begin{equation}\label{alpha0}
\alpha_0 = \frac{\int  A_0(\eta) \, \eta \dd \eta}{\int A^2_0(\eta) \, \eta \dd \eta}\per
\end{equation}

To investigate the accuracy of  \eqref{deriveA} we evaluate both sides  at $r=0$ for the simulation `L8a'. We obtain $A_0(\eta)$, with  normalization  $A_0(0)=1$, by numerical solution of eigenproblem \eqref{eigprob}. Using  this solution we  find that  left-hand side of  \eqref{deriveA} is  $\alpha_0 \phi_0 = 2.32$, where \eqref{alpha0} is used to calculate  $\alpha_0$; the constant $2.32$ is the dashed magneta line in figure  \ref{fig:RemoveConty}. Turning to the right-hand side of   \eqref{deriveA}, the blue sinusoidal curve  in figure \ref{fig:RemoveConty} is $|\phi(0,t)|$  computed using the baroclinic velocity fields of the Boussinesq simulation. The right-hand side of \eqref{deriveA} is obtained from the Boussinesq solution, resulting in  the red curve in figure \ref{fig:RemoveConty}. The time average of the red curve is $2.24$,  which is close to the prediction $\alpha_0 \phi_0=2.32$.

%\textcolor{green}{Is this an explanation of why the orange curve has small-amplitude wiggles? Old version: Because  the separation of time scales between $t$ and $\tau$ is  an approximation, $A_0$ maintains small variations in time. These variations are relatively small compared to the modulation of $|\phi|^2$ in figure \ref{fig:RemoveConty}. The average of the right hand side of (\ref{deriveA}) is $2.24$,  which is close to the theoretical prediction of the left hand side. }  

After gaining confidence in \eqref{deriveA}, we scale $A_0(r/a)$, which is computed by solving (\ref{eigprob}), by $\alpha_0 \phi_0 =2.32$ and compare it with the right-hand side of (\ref{deriveA}), averaged over $400$ inertial periods (about 3 eigenperiods) to remove the small variation in time that was discussed earlier. The results are shown in  figure \ref{fig:compare_eigfun}.  Despite many  approximations, the agreement between theory and simulation is remarkable. The tails of the two curves, however, display a noticeable difference stemming from the finite-domain effects -- point \textit{(ii)} above. Repeating the same process for several other simulations of table \ref{tab:list_sim}, we find similar agreement (not shown).

We emphasize that the joint excitation of the zeroth mode and continuum spectrum is necessary to observe the sub-inertial oscillations of the wave energy displayed in figures \ref{fig:HorizontalSlices} and \ref{fig:timeseries3points}. Because $\alpha_0$ and $A_0(r/a)$ are real, exciting solely the zeroth  mode results in a time-independent wave energy $|\phi_0|^2/2=\alpha_0^2 A_0^2(r/a)/2$.

\section{Nonlinear frequency shift}\label{sec:FreqShift}

\begin{figure}[h]
 \centerline{\includegraphics[width=19pc]{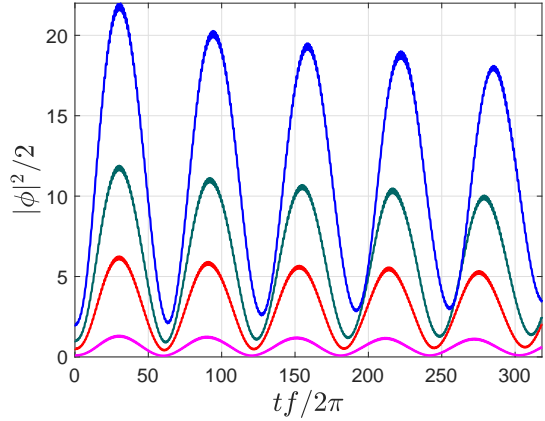}}
 \caption{Time series of wave kinetic energy $|\phi|^2/2$ at the centre of the vortex for fixed $\Ro$ and $\Bu$, and varying $E_0$. All the parameters in these simulations are the same as those in `L10C', expect for  $E_0$ which is set to $0.1$ for the magenta, $0.5$ for the red, $1.0$ for the green and $2.0$ for the blue curve.}\label{fig:DiffEo}
\end{figure}

According to \eqref{eigprob}, the oscillation of trapped modes depends only on $\lambda = \Ro/\Bu$. However, after fixing  these parameters,  we observe that the period of oscillations changes with the initial wave energy $E_0=\sqphin/2$: see figure \ref{fig:DiffEo}. To explain this frequency shift we have to go beyond the linear YBJ model. \cite{XV2015} include  the feedback of waves on the time-evolution of $\psi$ using a Generalised-Lagrangian-Mean (GLM) approach. \cite{wagner2015available} avoid GLM and instead present an alternative derivation using a multi-time expansion of the Eulerian equations of motion (see also \citep{wagner2016three}). The model can succinctly be written for a barotropic flow by adding a nonlinear wave-induced component, $q^W$, to the linear PV 

\begin{equation}\label{q_tot}
	q = \Dlt \psi + \underbrace{\frac{1}{f} \left[ \frac{1}{4} \Dlt \sqphi + \frac{i}{2} J(\phi^*,\phi) \right]}_{q^W}.
\end{equation}
The material conservation of $q$ together with the YBJ equation \eqref{WV_YBJ}  form a coupled model for  the joint evolution of $q$ and $\phi$, with $\psi$ obtained by inverting the Laplacian in \eqref{q_tot}   (see \cite{rocha2018stimulated} for the derivation). We emphasize that $\psi$ is the streamfunction associated with the Lagrangian mean flow; this is crucial for the interpretation 
of the model, including its energetics \citep{rocha2018stimulated,kafiabad2020wave}.

\begin{figure}
    \centerline{
     \includegraphics[width=19pc]{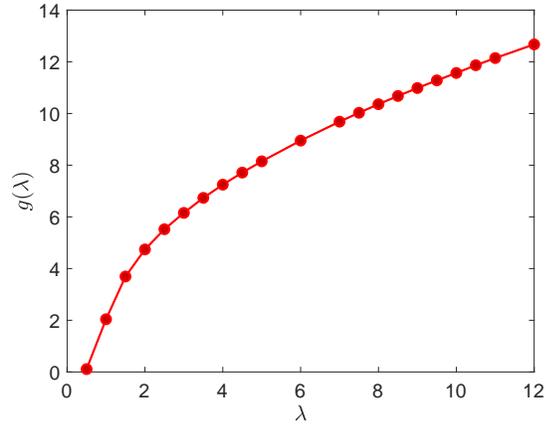}}
\caption{Numerical evaluation $g(\lambda)$  in \eqref{g_lambda}. The markers indicate numerical results and the smooth curve is an interpolant.}
\label{fig:glambda}
\end{figure}

The model simplifies dramatically when the wave and flow are axisymmetric. The potential vorticity (\ref{q_tot}) reduces to
\begin{align}
q(r) &= \frac{1}{r}  \frac{d}{d r} \left(  r  \frac{d \psi }{d r}  \right) + 
					 \frac{1}{4f} \frac{1}{r}  \frac{d}{d r} \left(  r  \frac{d \sqphi }{d r}  \right)\com \\
&=  \Dlt \left( \psi + \frac{ \sqphi}{4f} \right)\com
\end{align}
and its material conservation to the local invariance $\partial_t q =0$. 
For an initially uniform $\phi$, this gives
\begin{equation} \label{remarkable}
	q = \zeta_0 = \Dlt \psi_0  = \Dlt \left( \psi + \frac{\sqphi}{4f}  \right)\com
\end{equation}
\cite{kafiabad2020wave} confirm the validity of \eqref{remarkable} by comparing it against numerical solutions of Boussinesq equations.

%which can be integrated to obtain
%\begin{equation}\label{psi_phi}
%	\psi = \psi_0 - \frac{1}{4f} \sqphi \per
%\end{equation}
% \marginpar{\WRY{Why \eqref{psi_phi}???}} \marginpar{\WRY{Drop \eqref{psi_phi}???}}
% \marginpar{\WRY{Don't $$need$$ \eqref{psi_phi}???}}
%\WRY{The simple relation between the Lagrangian-averaged vertical vorticity and the back-rotated wave velocity in \eqref{remarkable} is investigated in detail in \cite{kafiabad2020wave},} where its validity is confirmed by comparing it against the high-resolution numerical solutions of Boussinesq equations.

Using  \eqref{remarkable}  to eliminate $\zeta = \lap \psi$  in  \eqref{WV_YBJ}  results in a closed nonlinear equation for $\phi$,
\begin{equation}\label{NLschrd}
	\frac{\partial \phi}{\partial t} + \frac{\ii}{2} \left( \Dlt \psi_0 -  \frac{\Dlt \sqphi  }{4f}  \right) \phi -  \frac{\ii}{2}  \hbar \ \Dlt \phi = 0 \per
\end{equation}
We are interested in the weakly nonlinear regime, when the cubic nonlinearity  $\Dlt \sqphi
 \, \phi/(4f)$ is small compared with the linear term $\Dlt \psi_0 \, \phi$, that is, when $\sqphi/(4f|\zetamin| a^2) \ll 1$. Based on this small parameter, we solve \eqref{NLschrd} by introducing the formal parameter $\varepsilon \ll 1$ and  rewriting \eqref{NLschrd} as 
\beq
\label{ShrdEqwEpsilon}
	\frac{\partial \phi}{\partial t} + \frac{i}{2} \left( \Dlt \psi_0 - \varepsilon   \frac{ \Dlt \sqphi}{4f}  \right) \phi -  \frac{i}{2}  \hbar \ \Dlt \phi = 0 \per
\eeq
We expand the back-rotated velocity and frequency according to
\begin{align}
\label{expansions}
	\phi &=  \phi_0 \alpha_0 A_0(r/a) e^{-\ii \omega t} + \varepsilon\  \tilde{\phi}(r,\varepsilon t) + \cdots, \\ 
	\omega &= \omega_0 + \varepsilon\ \tilde{\omega} + \cdots,
\end{align}
where $\omega_0$ and $A_0$ are the eigenvalue and eigenfunction of the zeroth trapped mode. 
Note that the leading order term in \eqref{expansions} varies on linear time scale $1/ \omega_0$ as well as slower time scale $\varepsilon t$, whereas the higher order terms only vary on the slower time scale. Computations detailed in Appendix B leads to the frequency shift
\begin{equation}\label{freqshift1}
	\tilde{\omega} =  \frac{\sqphin}{8f a^2} \ g(\lambda) \com
\end{equation}
where $g(\lambda)$ is the dimensionless function
\begin{equation}\label{g_lambda}
g(\lambda) = \cfrac{ \left( \mathop{\mathlarger{\int_0^\infty}} A_0 \eta d\eta \right)^2 \mathop{\mathlarger{\int_0^\infty}} \left( \cfrac{d }{d\eta} A_0^2 \right)^2 \eta d\eta }{\left( \mathop{\mathlarger{\int_0^\infty}} A_0^2 \eta d\eta \right)^3 }\per
\end{equation} 
The frequency shift $\tilde{\omega}$ is therefore quadratic in the wave amplitude $\phi$; in other words, $\tilde{\omega}$ scales linearly with the wave kinetic energy. The function $g(\lambda)$, computed from the numerical solution of the eigenproblem and shown in figure \ref{fig:glambda}, further shows that $\tilde{\omega}$  increases monotonically with $\lambda$; with small $\lambda$  the frequency shift is  less significant.

\begin{figure}
    \centering
     \includegraphics[width=19pc]{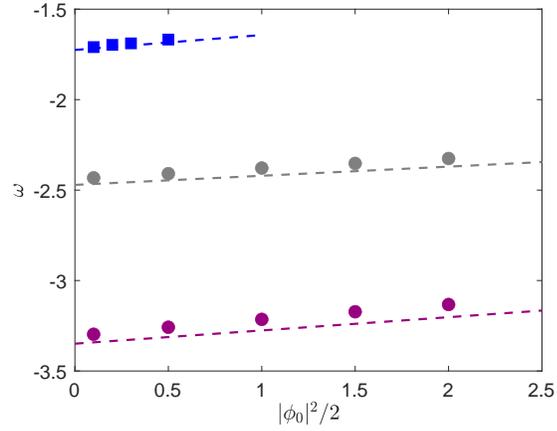}
\caption{Sub-inertial oscillation frequency as a function of initial wave energy: the predicted frequency  $\omega= \omega_0 + \tilde{\omega}$, where $\omega_0$ is the linear eigenfrequency and $\tilde{\omega}$ is a shift due to wave-feedback; $\tilde{\omega}$ is  calculated from \eqref{freqshift1} (dashed lines) and  compared with numerical estimates (circles and squares)  using  time series of wave energy at the vortex center. The parameters of simulation `L7', `L8c-R04' and `L10C' with varying $E_0 = \sqphin/2$ are used (the colour-coded markers are consistent with those in table \ref{tab:list_sim}).}
\label{fig:freqshift}
\end{figure}

In figure \ref{fig:freqshift}, the modified eigenfrequency that takes the wave-feedback into account, i.e.\ $\omega= \omega_0 + \tilde{\omega}$, is compared against the frequency of slow modulations in Boussinesq simulations. Three sets of simulations are considered where all the parameters are fixed within each set  while the initial wave energy $\sqphin/2$ is varied. These results show  good agreement between the nonlinear coupled model of \cite{XV2015} and the  Boussinesq simulations and confirm the validity of (\ref{freqshift1}). There is a small offset between the predicted and simulation frequencies of some sets, which is due to the issues \textit{(i)}-\textit{(iii)} discussed in the previous section. For `L8c-R04' (the blue markers) in figure  \ref{fig:freqshift} the analysis is limited to $E_0 < 0.5$: as discussed in the conclusion, for higher amplitude waves the vortex strongly interacts with the near inertial wave.

%: see figures \ref{fig:StrongIneracSlices} and \ref{fig:StrongInteracTseries}.

\section{Conclusion  and discussion}\label{sec:Conclusion}

\begin{figure*}[t]
 \centerline{\includegraphics[width=0.95\textwidth]{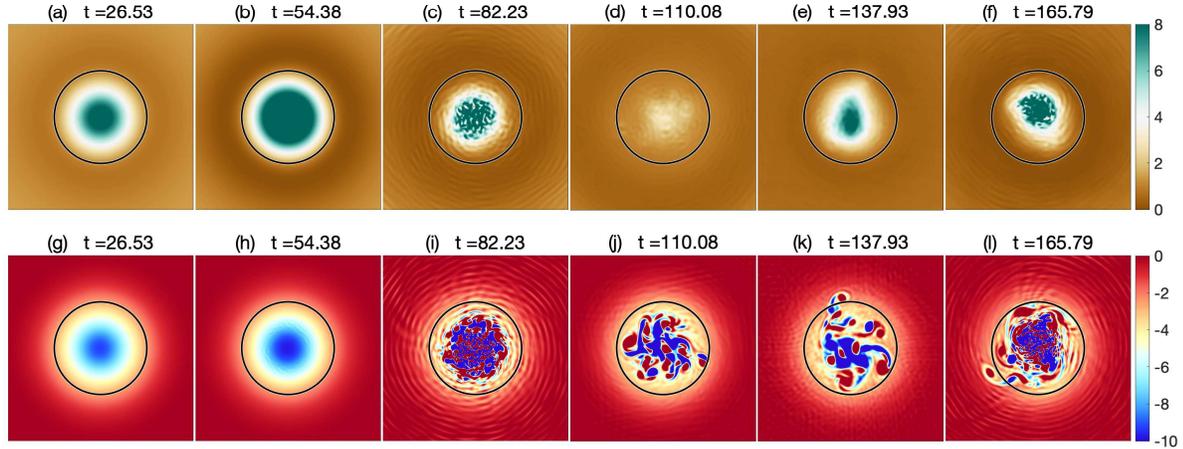}}
  \caption{Horizontal slices of the wave kinetic energy $(\uw^2 + \vw^2)/2$ (upper row) and the barotropic vertical vorticity (lower row). All the parameters in this simulation are the same as those in `L8c-R04', except that the initial wave energy is  increased  by a factor of thirty  to $E_0 = \sqphin/2 = 1.5$. Snapshots are taken at times indicated in inertial periods above each panel. }\label{fig:StrongIneracSlices}
\end{figure*}

\begin{figure}[h]
 \centerline{\includegraphics[width=19pc]{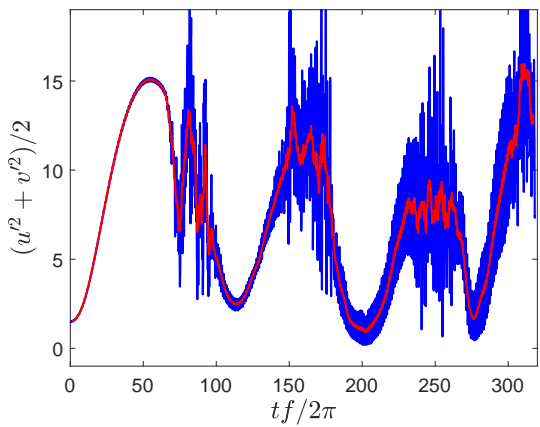}}
 \caption{Time series of wave kinetic energy $|\phi|^2/2$ at the centre of the vortex for the simulation shown in figure \ref{fig:StrongIneracSlices} in blue and its average over two inertial periods in red.}\label{fig:StrongInteracTseries}
\end{figure}

The linearized YBJ model of section \ref{sec:mathformulation}  focusses attention on the back-rotated velocity -- rather than  the radial velocity -- as the simplest variable characterizing  the trapped eigenmodes of an anticyclonic vortex. This is in immediate agreement with solutions of the Boussinesq equations: the top row of  figure \ref{fig:velocity_vec} shows that the trapped disturbance does not have a conspicuous radial velocity component.  Instead,  the back-rotated velocity is approximately  independent of azimuth. Thus the advective term,  $J(\psi,\phi)$ in  \eqref{WV_YBJ}, vanishes identically so that  near-inertial trapping results only from the $\ii \zeta \phi/2$ frequency shift.  In section \ref{sec:mathformulation}\ref{lowFreak} we show that as a consequence  the lowest possible frequency of the vortex  eigenmode -- the ``bottom of the discrete spectrum'' -- is $f + \zetamin/2$.

Exquisite excitation of a single pure eigenmode  produces a steady (time-independent)  axisymmetric pattern of wave kinetic energy density. But  generic forcing or initial conditions excites  all of the available discrete modes of the vortex, and also a continuum of untrapped disturbances.  Thus in the top row of figure  \ref{fig:HorizontalSlices}  we see that the  initial condition  in  \eqref{initWave} -- chosen to represent excitation by  an atmospheric storm of scale much larger than the vortex scale -- results in   a low-frequency pulsation of the kinetic energy density, corresponding to  the oscillations in the wave kinetic energy  time series of figure \ref{fig:timeseries3points}. This pulsation  is not a single pure eigenmode. In section \ref{sec:BoussinesqSim} we extracted the frequency of the sub-inertial  oscillation from figure  \ref{fig:timeseries3points} and showed that this modal  period is in good agreement with the predictions of the YBJ equation. 

In section \ref{sec:FreqShift} we go beyond the linear approximation and test the predictions of the phase-averaged model of \citet{XV2015}, \citet{wagner2016three} and \citet{rocha2018stimulated}. This model couples quasigeostrophic and YBJ models and accounts for the mean-flow change induced by wave feedback \citep[see also][]{kafiabad2020wave} through a  wave contributions to PV. We show that the wave feedback  leads to  frequency shift of the vortex eigenmode that is linearly proportional to the kinetic energy of the eigenmode. We find this frequency shift is in good quantitative agreement with Boussinesq results. This confirms the ability of the phase-averaged coupled model to represent NIW--mean flow interactions.

All  results in this work are in the regime of weak nonlinearity. But what happens if one hits the vortex with a very large initial disturbance? Figures \ref{fig:StrongIneracSlices} and \ref{fig:StrongInteracTseries} show the result of strongly perturbing a Gaussian vortex by increasing the amplitude of the initial condition $\phi_0$ in \eqref{initWave}. The initial development of this large  disturbance, up to about 55  inertial periods,   is similar to that of the weakly nonlinear problem shown in the top row of figure \ref{fig:HorizontalSlices}: the wave kinetic energy concentrates in the vortex core and the barotropic vorticity remains smooth. But after about 60 inertial periods  the core  concentration of wave kinetic energy  triggers an instability -- see figure \ref{fig:StrongInteracTseries}. The high-frequency bursts in figure \ref{fig:StrongInteracTseries} are  accompanied by  the formation of small spatial  scales in the vorticity field: in  figure \ref{fig:StrongIneracSlices} (j), (k) and (l), the main anticyclone  curdles  and small vortex dipoles circulate around its crumbled  remains. It  is impressive that a prominent  sub-inertial cycle persists  and   that there are episodes  of ``relaminarization'' coincident with the wave-energy minima in figure \ref{fig:StrongInteracTseries}. This  low-frequency  modulation of the instability is a persistent signature  of the vortex eigenmode that survives for over 300 inertial periods. 
There are open questions about the nature of the instability observed in figure \ref{fig:StrongIneracSlices} which we leave for future work.

%\WRY{ The strong interaction illustrated in figures \ref{fig:StrongIneracSlices} and \ref{fig:StrongInteracTseries}  is observed  if  the initial wave energy is sufficiently large. The signature of this regime is the periodic production   of  bursts of small-scale structure in both the near-inertial kinetic energy density and in barotropic vorticity. These bursts occur when the wave  kinetic energy density reaches  maximum amplitude in the vortex core. We do not know if these regime can be accessed by the wave-averaged system of \cite{XV2015}  and \cite{wagner2015available}. If, however, \eqref{remarkable} is valid  then we can rationalize these coincident bursts --- in fact we should check 
%$$
%\lap\left( \psi + \frac{ \sqphi }{4f}\right) \stackrel{?}{=} \lap\psi_0
%$$  
%for the run in  figures \ref{fig:StrongIneracSlices} and \ref{fig:StrongInteracTseries}!? Can we linearly combine the top and bottom rows to get rid of the small-scale obnoxiousness?}
%

%%

%%%%%%%%%%%%%%%%%%%%%%%%%%%%%%%%%%%%%%%%%%%%%%%%%%%%%%%%%%%%%%%%%%%%%
% DATA AVAILABILITY STATEMENT
%%%%%%%%%%%%%%%%%%%%%%%%%%%%%%%%%%%%%%%%%%%%%%%%%%%%%%%%%%%%%%%%%%%%%
% 
%
\datastatement

%%%%%%%%%%%%%%%%%%%%%%%%%%%%%%%%%%%%%%%%%%%%%%%%%%%%%%%%%%%%%%%%%%%%%
% ACKNOWLEDGMENTS
%%%%%%%%%%%%%%%%%%%%%%%%%%%%%%%%%%%%%%%%%%%%%%%%%%%%%%%%%%%%%%%%%%%%%
\acknowledgments
HAK and JV are supported by the UK Natural Environment Research Council grant NE/R006652/1. WRY is  supported by the National Science Foundation Award OCE-1657041. This work used the ARCHER UK National Supercomputing Service.

%%%%%%%%%%%%%%%%%%%%%%%%%%%%%%%%%%%%%%%%%%%%%%%%%%%%%%%%%%%%%%%%%%%%%
% APPENDIXES
%%%%%%%%%%%%%%%%%%%%%%%%%%%%%%%%%%%%%%%%%%%%%%%%%%%%%%%%%%%%%%%%%%%%%

%% If only one appendix, use
%%\appendix%

%% If more than one appendix, use \appendix[<letter>], e.g.,

\appendix[A] \label{App:WeakTrapping} 

\appendixtitle{The weak-trapping limit}

We solve the eigenvalue problem \eqref{eigprob} in the weak-trapping regime $\sigma \ll 1$ using matched asymptotics. 
In the outer region, $\eta \gg 1$, the Gaussian vorticity is exponentially small and can be neglected.  This leads to the outer approximation 
\beq
A(\eta) = q K_0\left(\sqrt{\sigma}\,  \eta\right)\com
\label{outer}
\eeq
where $K_0$ is the modified Bessel function and $q$ is undetermined constant. In an intermediate region where $\sqrt{\sigma}\,  \eta \ll 1$ and $\eta \gg 1$, the Bessel function $K_0$ is  approximated as
\beq
A(\eta) = - q \ln \eta - \tfrac{1}{2} q \ln \sigma  + q \big(\ln 2 - \gamE\big)+ \cdots, \label{match3}
\eeq
where $\gamE$ is Euler's constant and we have used the small-argument asymptotics of $K_0$.

In the inner region where $\eta = O(1)$, we use $\sigma \ll 1$ to reduce (\ref{eigprob}) to
\beq
\frac{\dd^2 A}{\dd \eta^2} + \frac{1}{\eta}\frac{\dd A}{\dd \eta} + \lambda \ee^{-\eta^2} A= 0\per
\label{inner}
\eeq
We select the bounded  solution  as $\eta \to 0$ by imposing
\beq
A(0) = 1 \quad \textrm{and} \quad A'(0)=0\per
\label{innerBC}
\eeq
Equations \eqref{inner} and \eqref{innerBC}  define an initial-value problem which --  except of the zeroth mode in  \eqref{zerothA} below -- must  be solved numerically. For $\eta \gg 1$, the solution has the asymptotics
\beq
A(\eta) \sim \alpha(\lambda) \ln \eta + \beta(\lambda) + o(1), 
\label{eq:alphabeta} 
\eeq
with  $\alpha(\lambda)$ and $\beta(\lambda)$  determined from the numerical solution. Matching \eqref{eq:alphabeta} with \eqref{match3} results in 
\beq
\frac{\beta(\lambda)}{\alpha(\lambda)} = \tfrac{1}{2} \ln \sigma + \gamE - \ln 2\per
\label{aaa}
\eeq
This is an equation for $\sigma$ that is valid only for $\beta/\alpha < 0$ and $|\beta/\alpha| \gg 1$  so that $\sigma \ll 1$ as assumed. Equation \eqref{aaa} identifies the zeros $\lambda_n, \ n=0,\, 1,\cdots$ of the function $\alpha(\lambda)$ as the values of $\lambda$ at which new branches of the dispersion relation appear. Note that $\lambda_0=0$ corresponds to the zeroth mode; this eigensolution exists even for very weak vortices.

\begin{figure}
\centerline{\includegraphics[width=19pc]{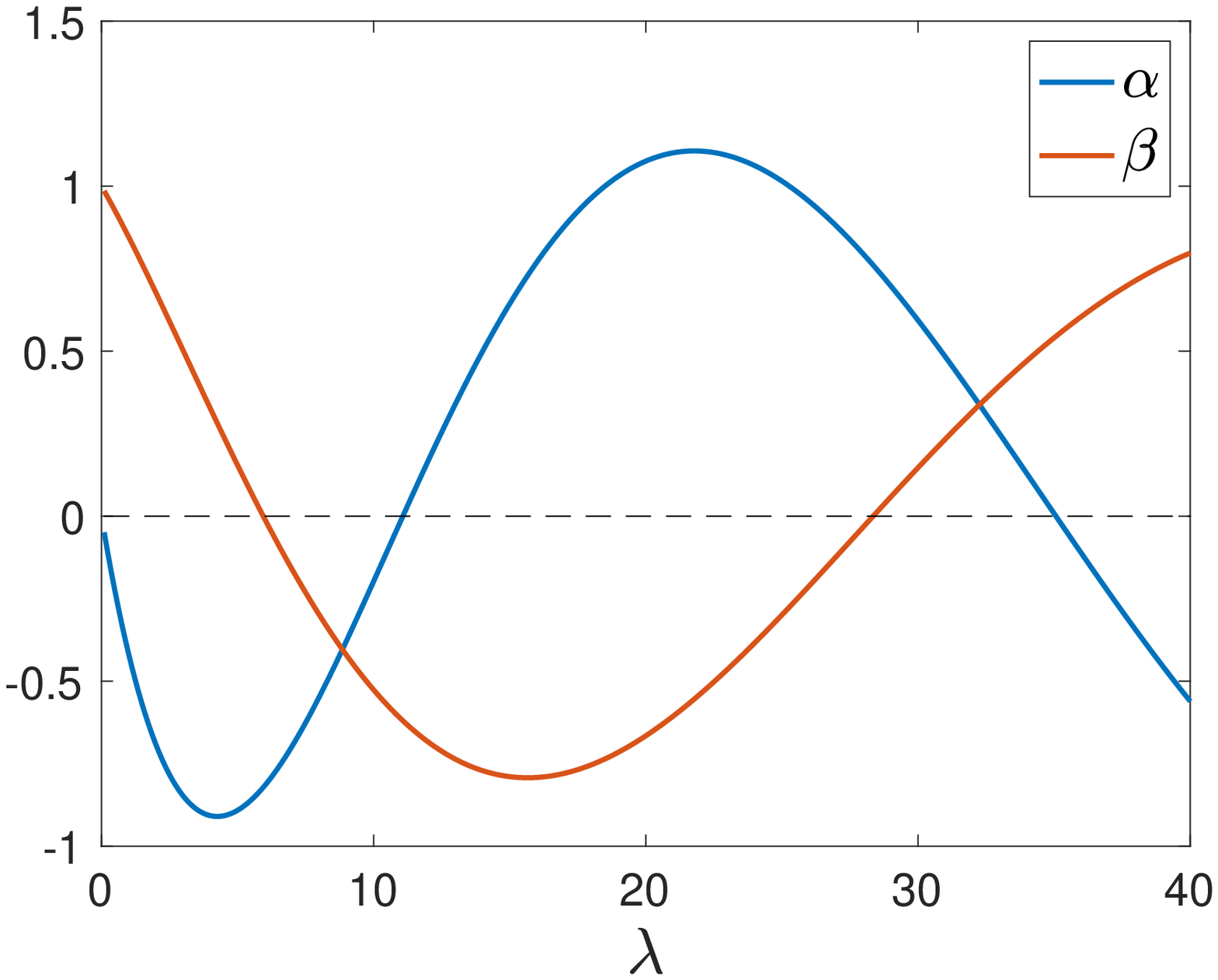}}
\appendcaption{B1}{Functions $\alpha(\lambda)$ and $\beta(\lambda)$ derived by solving (\ref{inner}) numerically.}
\label{fig:alphabeta}
\end{figure}

We have computed $\alpha$ and $\beta$ numerically for  $0<\lambda\leq 40$ and show the results in figure \ref{fig:alphabeta}. The first bifurcation values of $\lambda$ are found to be $\lambda_0 = 0$, $\lambda_1 \approx 11.1$ and $\lambda_1 \approx 35.1$.  
In view of the sign of $\beta(\lambda_n)$, new branches appear for $\lambda > \lambda_n$. Approximating the left-hand side of \eqref{aaa} near $\lambda_n$ and solving for $\sigma$ leads to the dispersion relations
\beq\label{asympt4lambda_n}
\sigma \sim e^{2 (\ln 2 - \gamE) - c_n / (\lambda - \lambda_n)} \quad \textrm{as} \ \ \lambda \to \lambda_n^+,
\eeq
with $c_n = - 2 \beta(\lambda_n)/\alpha'(\lambda_n)>0$. 

We can obtain a fully analytic form for the $n=0$ branch (the zeroth mode), with $\lambda \approx \lambda_0 = 0$ by solving \eqref{inner} asymptotically for $\lambda \ll 1$. A straightforward  expansion in powers of $\lambda$ gives
\begin{align}
A(\eta)&= 1 - \frac{\lambda}{4} \int_0^{\eta^2} \frac{e^{-x} - 1}{x} \, \dd x + O(\lambda^2) \, ,\\
&= 1 - \tfrac{1}{4} \lambda \left(E_1(\eta^2) + 2 \ln \eta \right) + O(\lambda^2)\com \label{zerothA}
\end{align} 
where $E_1$ is the exponential integral. Noting that $E_1(\eta) \to 0$ as $\eta \to \infty$, we find from \eqref{eq:alphabeta} that as $\lambda \to 0$, $\alpha(\lambda) \sim - \lambda/2$ and $\beta(\lambda) \sim 1$; this results  in $c_0 = 4$ and the zeroth-mode dispersion relation in \eqref{eq:smallSigma}.
For the $n=1$ branch, we find numerically $\alpha'(\lambda_1) \approx 0.18$ and $\beta(\lambda_1) \approx - 0.62$; hence $c_1 \approx 6.8$.

 \appendix[B] 
\appendixtitle{Frequency shift due to wave feedback}

Substituting \eqref{expansions} into (\ref{ShrdEqwEpsilon}),  and keeping the terms at order $\varepsilon^0$, leads to 
\begin{equation}\label{order0}
	- \omega_0 A_0 + \frac{1}{2} \Dlt \psi_0 A_0 - \frac{1}{2} \hbar \Dlt A_0 = 0,
\end{equation}
which is the dimensional form of (\ref{eigprob}) for $A_0$ and $\omega_0$. Based on this equation, we introduce  the self-adjoint operator $\mathcal{L}$
\begin{equation}
	\mathcal{L}  = - \omega_0+ \frac{1}{2} \Dlt \psi_0 - \frac{1}{2} \hbar \Dlt. 
\end{equation}
The terms at the next order form the following equation
\beq\label{odrer1}
	- \tilde{\omega} A_0 + \frac{\mathcal{L} \tilde{\phi}}{\alpha_0 \phi_0}  - \frac{\alpha_0^2 \sqphin}{8f} \Dlt A_0^2 \ A_0  = 0,
\eeq	
which can be multiplied by $A_0$ and then integrated to obtain
\begin{align}\label{Xxx}
	-  \tilde{\omega} \int A_0^2 \,  r \dd r + \int  (\mathcal{L} \tilde{A}) A_0 \, r  \dd r 
	   - \frac{1}{8f} \int A_0^2 \ \Dlt A_0^2  \, r \, \dd r  = 0
\end{align}
(All  integral run  from $r=0$ to  $\infty$). Because of the self-adjoint property of $\mathcal{L}$ and (\ref{order0}),  
\beq\label{X5}
	\int  (\mathcal{L} \tilde{\phi}) A_0 r \, \dd r = \int  \tilde{\phi} (\mathcal{L}  A_0) r\,  \dd r = 0. 
\eeq
The last integral in (\ref{Xxx}) can also be simplified after integration by parts
\begin{align}\label{X6}
	\int A_0^2 \Dlt A_0^2  \ r \, \dd r &= \int A_0^2 \frac{\dd}{\dd r} \left(r \frac{\dd}{\dd r} A_0^2 \right)  \,\dd r \nonumber \\
	&= - \int \left( \frac{\dd}{\dd r} A_0^2 \right)^2  \ r \, \dd r 
\end{align}
Using (\ref{X5}) and (\ref{X6}), (\ref{Xxx}) reduces to the following expression after rewriting the integrals in terms of the  dimensionless coordinate $\eta=r/a$
\begin{equation}
	\tilde{\omega} = \frac{\sqphin}{8f a^2}  \cfrac{ \left( \mathop{\mathlarger{\int}} A_0 \eta \dd\eta \right)^2 \mathop{\mathlarger{\int}} \left( \cfrac{\dd }{\dd\eta} A_0^2 \right)^2 \eta \dd\eta }{\left( \mathop{\mathlarger{\int}} A_0^2 \eta \dd\eta \right)^3 },
\end{equation}
where we used (\ref{alpha0}) to substitute for $\alpha_0$.

\bibliographystyle{ametsoc2014}
\bibliography{references.bib}

\end{document}